\documentclass[research]{fcs}
\pdfoutput=1
\usepackage[utf8]{inputenc}
\usepackage[T1]{fontenc}
\usepackage{subcaption}
\usepackage{enumitem}
\usepackage{multirow} 
\usepackage{booktabs}
\usepackage{tikz} 
\usepackage{tcolorbox}
\usepackage{makecell}
\usepackage{xcolor} 
\usepackage{amsmath} 
\usepackage{graphicx} 
\usepackage{adjustbox} 
\usepackage{tabularx} 
\usepackage{subcaption}
\usepackage{fancyhdr}
\usepackage{lastpage} 
\usepackage{enumitem}

\title{Self-Supervised Representation Learning with ID-Content Modality Alignment for Sequential Recommendation}
\author[1]{Donglin Zhou}
\author*[1]{Weike Pan}
\author[2]{Zhong Ming}
\address[1]{College of Computer Science and Software Engineering, Shenzhen University, Shenzhen 518060, China}
\address[2]{Guangdong Laboratory of Artificial Intelligence and Digital Economy (SZ), Shenzhen 518123, China}
\fcssetup{
  received       = {month dd, yyyy},
  accepted       = {month dd, yyyy},
  corr-email     = {panweike@szu.edu.cn},
}

\begin{abstract}
Sequential recommendation (SR) models often capture user preferences based on the historically interacted item IDs, which usually obtain sub-optimal performance when the interaction history is limited. Content-based sequential recommendation has recently emerged as a promising direction that exploits items' textual and visual features to enhance preference learning. However, there are still three key challenges: (i) how to reduce the semantic gap between different content modality representations; (ii) how to jointly model user behavior preferences and content preferences; and (iii) how to design an effective training strategy to align ID representations and content representations. To address these challenges, we propose a novel model, \underline{s}elf-supervised representation learning with \underline{I}D-\underline{C}ontent modality alignment, named SICSRec. 
Firstly, we propose a LLM-driven sample construction method and develop a supervised fine-tuning approach to align item-level modality representations. Secondly, we design a novel Transformer-based sequential model, where an ID-modality sequence encoder captures user behavior preferences, a content-modality sequence encoder learns user content preferences, and a mix-modality sequence decoder grasps the intrinsic relationship between these two types of preferences. Thirdly, we propose a two-step training strategy with a content-aware contrastive learning task to align modality representations and ID representations, which decouples the training process of content modality dependency and item collaborative dependency. Extensive experiments conducted on four public video streaming datasets demonstrate our SICSRec outperforms the state-of-the-art ID-modality sequential recommenders and content-modality sequential recommenders by 8.04\% on NDCG@5 and 6.62\% on NDCD@10 on average, respectively.
\end{abstract}
\keywords{Sequential recommendation, supervised fine-tuning, representation alignment.}

\begin{document}

\section{Introduction}
Sequential recommendation aims to model user preferences based on their historical behaviors (i.e., click, purchase, etc.) and predict the next item for the user. Existing methods utilize a unique item ID to represent each item, and design different neural networks to model user-item interactions (e.g., RNN \cite{GRU4Rec}, Transformer \cite{sasrec, bert4rec}, MLP \cite{BMLP} and GNN \cite{srgnn, BA-GNN}). This ID modality-based sequential modeling paradigm essentially learns item-to-item collaborative dependency from a statistical perspective. Previous works have achieved great success in the last decade, but often lead to sub-optimal performance and poor generalization ability when user-item interactions are few \cite{tailrec}. 

Recently, content-modality based recommendation becomes a promising direction that exploits the item content modality information (e.g., text and image) to enhance user preference learning \cite{SPACE, SSL_MMRec}. These content modality data widely exist on online platforms such as YouTube, TikTok, and BiliBili, which contain rich semantic information about the items. Furthermore, modeling user content modality information gives a comprehensive view to understand user preferences \cite{dimo, Showcases}. For example, in a fashion recommendation scenario, a user may be attracted by a stylish clothing image or a promotional description \cite{Fashion_rec}. The decision of users to click or not within an online website is usually impacted by the visual characteristics \cite{COURIER}. 

Existing content modality-based sequential recommendation methods model the semantic similarity between different items to enhance the item representations, and combine the users' content dependency and item collaborative dependency to improve the recommendation performance \cite{unsirec, vqrec, MISSRec}. Despite the advanced progress, there are still challenges.

(1) \textit{How to reduce the semantic gap between different content modality information and obtain a unified item content representations?} The item content modality representations come from a pre-trained language embedding space or a vision embedding space. Due to differences in training data and optimization objectives, there is an inherent semantic gap between the text modality representations and image modality representations, even when they represent the same item \cite{MMSSL, AlignRec, mindgap}.

(2) \textit{How to acquire accurate user preferences by jointly modeling content sequences and ID sequences?} In our methods, we mainly consider two preferences, i.e., behavior preferences and content preferences. User behavior preferences tend to reflect the item-to-item collaborative dependency by modeling ID sequences. However, the content representations of a corresponding item sequence lacks an understanding of the personal intents. It thus needs to explicitly learn user intents from both the content sequences and ID sequences for better user preference learning. 

(3) \textit{How to design an effective training strategy to align ID modality representations and content modality representations?} The item ID modality is the most commonly used modality, which serves as a fine-grained numerical feature. The item content modality comes from a pre-trained embedding space, and provides more coarse-grained semantic information. Due to the heterogeneity of these two modalities, end-to-end training of a recommender that combines content representations and ID representations may lead to unstable performance caused by representation interference.

To tackle these issues, we propose a novel self-supervised representation learning framework (i.e., SICSRec) with ID-content modality alignment for sequential recommendation. In this paper, we mainly consider two types of modality information, i.e., item ID, and item content (i.e., text and image). 

For the first challenge, we propose a novel content modality semantic alignment module to reduce the semantic gap between different modalities. We first propose a novel LLM-driven sample construction method, which leverages LLM as a semantic discriminator to select the most similar item-content modality pairs from users' interaction sequences. Then, we design a supervised fine-tuning approach to jointly tune the text and image encoders, which facilitates item-level modality representation alignment and obtains representations that are better suitable for downstream recommendation tasks.

For the second challenge, we first adopt an L2 normalization to transfer the multi-content representations into a unified content representation space. Then we develop a novel Transformer-based encoder-decoder model, where an ID-modality sequence encoder captures user behavior preferences from the item-ID sequence, a content-modality sequence encoder learns user content preferences from the item-content sequence, and a mix-modality sequence decoder grasps the intrinsic relationship between these two types of preferences. We aggregate these three outputs as the final user preferences.

For the third challenge, we adopt a two-step training strategy to train our model, which decouples the training process of content dependency and item collaborative dependency. Firstly, we pre-train an ID modality sequence encoder with a standard cross-entropy loss, and then fix its weights. Then, we propose content-aware contrastive learning as an auxiliary loss, which aligns user preferences with content-modality representations. We utilize a low-rank adaptation layer to extract user behavior preferences from the ID-modality encoder, and post-train other components.

In the experiments, we compare our SICSRec with eleven competitive baselines including ID modality-based sequential recommenders and content modality-based sequential recommenders on four public video streaming datasets. Our SICSRec achieves significant improvement in ranking-oriented evaluation metrics. Moreover, we conduct extensive ablation studies to validate our model's components. The contributions are summarized as follows:
\begin{itemize}
\item [(1)] We propose a novel content-modality semantic alignment method, reducing the semantic gap between different content-modality representations. It reveals that using LLM for data construction and supervised fine-tuning for the content encoder is a promising way to enhance item-level content representations.
\item [(2)] We design a Transformer-based encoder-decoder architecture to model user behaviors, content preferences, and their inherent relationships. We further introduce a novel content-aware contrastive learning task and an effective two-step training strategy to combine content dependency and item collaborative dependency, facilitating efficient representation learning.
\item [(3)] Extensive experiments on four public datasets show that our SICSRec outperforms the state-of-the-art baselines. Additionally, the ablation study highlights the effectiveness of each key component and the robustness of performance.
\end{itemize}

\section{Related Work}
\subsection{ID-based Sequential Recommendation}
Classical sequential recommendation methods utilize unique item IDs to represent each item and design different neural networks to model a user's long-term and short-term preferences. Previous works have introduced RNN (e.g., GRU4Rec \cite{GRU4Rec}), Transformer (e.g., SASRec \cite{sasrec}, BERT4Rec \cite{bert4rec}, Transformer4Rec \cite{transformer4rec}, RETR \cite{pathway}), MLP (e.g., FMLP-Rec \cite{filter-mlp}, MMMLP \cite{MMMLP} and BMLP \cite{BMLP}) and GNN (e.g., SR-GNN \cite{srgnn} and BA-GNN \cite{BA-GNN}) to sequential modeling. 

Further works leverage side information, such as item category or price as prior knowledge, and design different fusion networks to enhance item representations \cite{nova, beyond}. For example, Cafe \cite{cafe} explicitly learns coarse-grained user intents for item category sequences. DIF \cite{DIF} proposes a decoupled fusion method to adaptively fuse side information and item representations. Furthermore, MSSR \cite{MSSR} designs a multi-sequence integrated attention layer and a user representation alignment module to optimize representation learning. 

Self-supervised learning is utilized to grasp supervision signals from user interaction sequences or attribute sequences to enhance user preference learning \cite{intentcl, duorec}. For example, SelfGNN \cite{selfgnn} encodes short-term collaborative relationships via graph neural networks and captures stable user representations via self-augmented learning. S3Rec \cite{S3} proposes four auxiliary self-supervised objectives to learn the intrinsic data correlation and enhance user representation learning. DuoRec \cite{duorec} focuses on the representation degeneration issue and proposes a contrastive regularization term to alleviate it. DCRec \cite{dcrec2023} adopts a debiased contrastive learning paradigm to capture item-level and user-level dependencies. Furthermore, Poisoning-SSL \cite{Poisoning} explores the feasibility of the poisoning attacks on self-supervised learning methods and the weaknesses of these methods.

The modeling paradigm of the above methods can be called item ID modality-based sequential recommendation, which represents each user or item with ID representations, and essentially learns the item-item collaborative relationship from a statistical perspective. However, despite advanced progress, the item ID-based sequential recommendation still struggles with data sparsity and cold-start issues, leading to sub-optimal performance and poor generalization ability when user interactions are few. 

\subsection{Content-based Sequential Recommendation}
Content-based sequential recommendation aims to utilize item content information (e.g., item image and item text) to enhance item representations \cite{dimo, multifusion, IDGenRec}. This modeling paradigm can be categorized into two main branches:  content-centric SR methods and content-enhanced SR methods. 

Inspired by the success of pre-trained models, content-centric SR methods focus on using text or image modality representations to replace the item-ID modality representations and directly conduct end-to-end recommendation. For example, MoRec trains the modality encoder and recommender jointly with end-to-end training, which achieves similar performance compared to the ID modality-based recommender in some scenarios \cite{IDvs.MoRec, transfer}. Recformer \cite{recformer} models item text features as language representations and trains Longformer to understand recommendation tasks. TASTE \cite{textmatch} uses T5 as the backbone,  represents items and users with text, and predicts the next item for each user based on the relevance of the text representations. However, end-to-end training for content modality-based recommenders is costly, which may not meet real-time inference requirements in personalized recommendation.

The content-enhanced SR methods utilize a pre-trained modality model as a feature encoder to obtain the item-level content representations, and then combine them with the ID representations to enhance the recommendation performance. This modeling paradigm is effective and efficient to deploy in industry. For example, some works utilize pre-trained BERT to extract side features from reviews to enhance item representations \cite{U-bert, cascaded}. UniSRec \cite{unsirec} learns universal item representations from associated description text of items, and introduces two contrastive pre-training tasks to build transferable recommendation. VQ-Rec \cite{vqrec} maps item text embedding into multiple code embedding, and designs a differentiable permutation-based network for recommendation. MISSRec extends UniSRec to a multi-modal learning framework, which jointly models image and text preferences \cite{MISSRec}. Moreover, MSRec \cite{moe4rec} proposes a mixture-of-experts (MoE) fusion network for multi-modal information fusion. MML \cite{MMLRec} designs a group of multimodal meta-learners, each learns the corresponding kind of modality information, and fuses them in the prediction layer. M5 \cite{m5} learns content graph embeddings from a metagraph, and combines the ID embeddings in multi-interest extraction layer. 

These works design different deep networks, which combines the ID modality representations and content modality representations to enhance user preference learning. However, these works rarely consider the item-level semantic gap between different content modality representations and the heterogeneity of content modality and ID modality, which may lead to unstable performance and difficulty in model convergence.

\section{Methodology}
\subsection{Problem Definition}
\label{define}
For each $u \in \mathcal{U}$, we define $S_{id}=\{i_1,i_2,..,i_n\}$ as the item-ID sequence, $S_{text} =\{t_1,t_2,...,t_n\}$ as the item-text sequence, and $S_{ima} =\{g_1, g_2, ..., g_n\}$ as the item-image sequence, where $n$ is the fixed length of a sequence. These three sequences are sorted in chronological order. For each item $i_k$ at time step $k$, there is a corresponding content pair $(t_i, g_i)$, where $t_i$ is the item text and $g_i$ is the item image. Our goal is to exploit these three sequences to predict the user’s next preferred item $i_{n+1}$,  which can be formulated as follows,
\begin{equation}
\arg\max(i_{n+1}|S_{id}, S_{text}, S_{ima}).
\end{equation}

\subsection{Overview of Our SICSRec}
\begin{figure*}[ht]
    \centering
    \includegraphics[width=\linewidth]{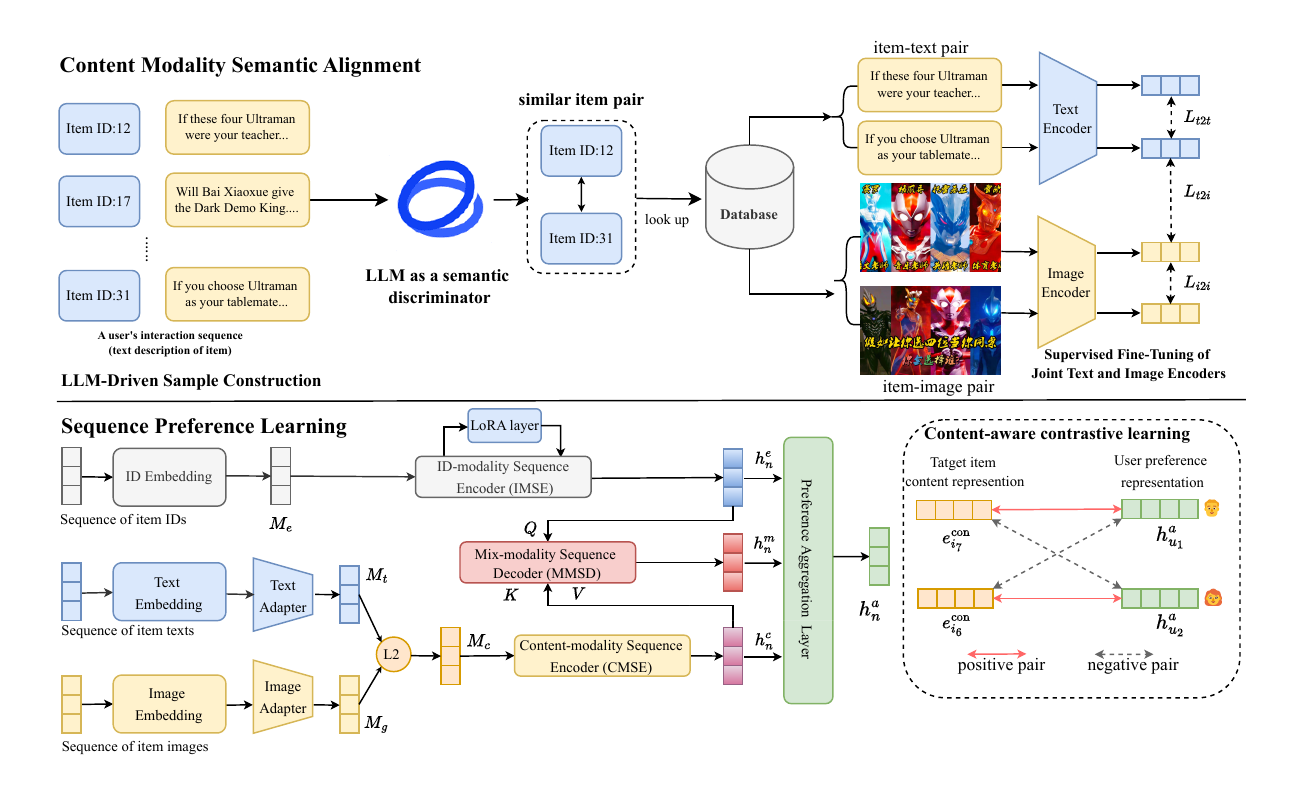}
    \caption{The architecture of our SICSRec, including (i) content modality semantic alignment in Section \ref{item_level_modality_alingment} and (ii) sequence preference learning in Section \ref{Sequence-level preference modeling}. We propose a two-step training strategy with a content-aware contrastive learning task, which is described in Section \ref{training_strategy}.}
    \label{fig:overall}
\end{figure*}

The overall framework of our SICSRec is illustrated in Figure~\ref{fig:overall}, consisting of three parts, i.e., (i) content modality semantic alignment in Section \ref{item_level_modality_alingment}, 
(ii) sequence preference learning in Section \ref{Sequence-level preference modeling}, and (iii) a two-step training strategy with content-aware contrastive learning task in Section \ref{training_strategy}.  In the first part, we design an LLM-driven sample construction method and a supervised fine-tuning method of joint text encoder and image encoder to achieve item-level modality representation alignment. In the second part, we propose a Transformer-based encoder-decoder model for user preference learning. In the third part, we design a content-aware contrastive learning task to align content representations and ID representations and adopt a two-step training strategy to train our model. The important notations and their explanations are listed in Table~\ref{tab:Notations}.

\begin{table}[htbp]
  \caption{Notations and their explanations.}
  \label{tab:Notations}
  \begin{tabular}{cp{6cm}}
    \toprule
   Notation&Description\\
    \midrule
    $U$ & The user set \\
    $I$ & The item set \\
    $S_{id}$ & The item-ID sequence \\
    $S_{text}$ & The item-text sequence \\
    $S_{ima} $ & The item-image sequence \\
    $M_e \in \mathbb{R}^{|I|\times d} $ & The input item-ID embedding matrix \\
    $M_t \in \mathbb{R}^{|I|\times d} $ & The input item-text embedding matrix \\
    $M_g \in \mathbb{R}^{|I|\times d} $ & The input item-image embedding matrix \\
    $M_c \in \mathbb{R}^{|I|\times d}$ & The input item-content embedding matrix \\ 
    $n$ & the maximum length of a sequence\\
    $d$ & the latent vector dimension \\
  \bottomrule
\end{tabular}
\end{table}

\subsection{Content Modality Semantic Alignment}
\label{item_level_modality_alingment}
Directly using the pre-trained content representations in the recommendation model may lead to sub-optimal performance. There are two main reasons: first, the item content representations (i.e., text and image) come from a pre-trained language embedding space and a vision embedding space. Due to the inconsistency in training data and objectives, there is an item-level semantic gap between text and image representations for the same item. Secondly, the original content representations lack an understanding of the item-item collaborative information for a specific recommendation scenario because the content encoders do not exploit the user-item interaction history.

Some previous works \cite{jinaclip, enhance_taobao} show that fine-tuning a content encoder for a domain-specific data can effectively improve the content representations in downstream tasks. Large language model has powerful semantic discriminative capabilities, which have achieved significant success across a range of tasks, such as text generation and classification. Inspired by the sample efficiency of LLM-enhanced recommender systems\cite{llmsamole}, in this paper, we propose an LLM-driven sample construction method, which uses an LLM to select some most similar content pairs from a user's historically interacted items. Then, we design an effective supervised fine-tuning method for item-level modality representation alignment, which jointly tunes the text encoder and the image encoder based on the LLM-selected data. After that, we utilize these tuned content encoders to obtain high-quality content embeddings, and combine them with ID embeddings in the downstream recommendation tasks. 

\subsubsection{LLM-driven Sample Construction}
\label{LLM-Driven Sample Construction}
We select a powerful large language model as a semantic discriminator to select semantically similar item pairs. Specifically, we first construct task-specific prompts based on the user-item interaction history. The prompt contains three parts: instruction, input, and output guidance. The instruction part defines a specific task and instructs the LLM to engage in role-playing. The input part is the target item text and the candidate items. The output guidance requires the LLM to generate content in the correct format. In the input part, given a user-item interaction history, we choose the title of the final item in the user-item interaction sequence as the target item text, and the same user's previously interacted items' titles as the candidates. 

We design this prompt to activate the reasoning ability and open-world knowledge of LLM. The LLM selects the semantically most similar item from the candidate ones according to the text semantics of the target item and outputs the corresponding item-ID pair. Notice that LLM evaluates the semantic similarity between items based on raw text analysis, instead of on embedding-based retrieval. The selected item pair can be considered as having both semantic similarity and a collaborative co-occurrence pattern. 

The prompt template is as follows.
\begin{tcolorbox}[colframe=black,colback=white,title=Semantic Sample Construction Prompt]
\textbf{<Instruction>:} You are a video similarity evaluation assistant. I will provide you with a target video title and a list of candidate video titles. Please help me find the most similar video title from the candidate list to the target video. \\
\textbf{<Input>:} The target video is \textit{<target\_itemID>} - \textit{<target\_item\_title>}, and the candidate videos are: \textit{<candidate\_item\_descriptions>}. \\
\textbf{<Output Guidance>:} Please find the most similar video title from the candidates and output the corresponding item-ID pair in the following format: \textit{<target\_itemID>-<similar\_itemID>}. \\
If there are no similar videos, output (-1,-1) directly. Please ensure the format is correct; any other format will be considered invalid.
\end{tcolorbox}

In this template, the special token \textit{<target\_itemID>}-\textit{<target\_item\_title>} is replaced by the target item ID and its title. 

The special token \textit{<candidate\_item\_descriptions>} consists of several tokens like \textit{<candidate\_itemID>-<candidate\_item\_title>}, which is constructed from the same user's historically interacted items. We feed the complete prompt to an LLM and obtain a semantically most similar item-ID pair.

\subsubsection{Supervised Fine-tuning of Joint Text and Image Encoders}
\label{Supervised Fine-Tuning of Joint Text and Image Encoders}
Supervised fine-tuning facilitates a content encoder to generate better semantic representations for similar items in a downstream task and also align different content representations for a same item. Therefore, we define three alignment tasks, including text-to-text (t2t) alignment, image-to-image (i2i) alignment, and text-to-image (t2i) alignment. 

Text-to-text (t2t) alignment: Given a batch of text pairs, we consider each text pair as a positive sample and target items' texts from other text pairs in the batch as negative samples. We pull text representations of positive samples and push apart the text representations of negative samples. We adopt the classical InfoNCE \cite{self-item} loss, 
\begin{equation}
\begin{aligned}
\mathcal{L}_{t2t} &= -\frac{1}{N} \sum_{i=1}^{N} \log \left( \frac{\exp\left( B_{ii}^{t2t} \right)}{\exp\left( B_{ii}^{t2t} \right) + \sum_{j \neq i} \exp\left( B_{ij}^{t2t} \right)} \right), \\
    B_{ii}^{t2t} &=  \textit{sim}(e_{i_p}^{text}, e_{i_q}^{text}) / \tau, \\
    B_{ij}^{t2t} &= \textit{sim}(e_{i_p}^{text}, e_{j_q}^{text}) / \tau, \\
\end{aligned}
\end{equation}
where $N$ is the batch size, $sim$ is the vector inner product operation, and $\tau \in[0,1]$ is the temperature parameter. Note that $e_{i_p}^{text} \in \mathbb{R}^{1 \times d}$ and $e_{i_q}^{text} \in \mathbb{R}^{1 \times d}$ represent the $i$-th positive text representation pair. $B_{ii}^{t2t}$ is the representation similarity of the positive text pair and $ \sum_{i \neq j} B_{ij}^{t2t}$  is the sum of negative-instance similarities. Similarly, we have a loss for the image-to-image alignment $\mathcal{L}_{i2i}$.

Text-to-image (t2i) alignment: Inspired by the CLIP loss \cite{clip}, we design a text-to-image (t2i) alignment task. We use the text and image representations of a same item as a positive sample pair, and those of different items as negative sample pairs. We again use the InfoNCE \cite{self-item} loss,
\begin{equation}
\begin{aligned}
\mathcal{L}_{t2i} &= -\frac{1}{N} \sum_{i=1}^{N} \log \left( \frac{\exp\left( B_{ii}^{t2i} \right)}{\exp\left( B_{ii}^{t2i} \right) + \sum_{j 
eq i} \exp\left( B_{ij}^{t2i} \right)} \right), \\
 B_{ii}^{t2i} &= \textit{sim}(e_i^{text}, e_i^{ima}) / \tau, \\
 B_{ij}^{t2i} &= \textit{sim}(e_i^{text}, e_j^{ima}), / \tau
\end{aligned}
\end{equation}
where $B_{ii}^{t2i}$ is the representation similarity of the text representations and image representations of a same item $i$, and $\sum_{i \neq j} B_{ij}^{t2i}$ is the sum of similarities of negative sample pairs. 

Finally, we obtain the final supervised fine-tuning loss, 
\begin{equation}\label{sft}
\mathcal{L}_{SFT} = \mathcal{L}_{t2t} + \mathcal{L}_{i2i} + \mathcal{L}_{t2i}.
\end{equation}
We use $\mathcal{L}_{SFT}$ to jointly fine-tune the text encoder and image encoder for each downstream task, which achieves item-level modality representation alignment. Then, we use the tuned content encoder to encode each item's text and item image as semantic representations. These semantic representations would engage in sequence preference learning.

\subsection{Sequence Preference Learning}
\label{Sequence-level preference modeling}
User dynamic preferences can be captured by their previous interaction history, as well as by the content modality (e.g., title or image) of the items. Therefore, we consider two types of preferences in our model, i.e., behavior preferences on item IDs and content preferences. 

Firstly, we construct an item-ID embedding matrix and an item-content embedding matrix via an embedding layer. Secondly, we unified the image representations and text representations as unified content representations via L2-normalization. Then, we develop a Transformer-based encoder-decoder sequence model, where an ID-modality sequence encoder captures user behavior preferences, a content-modality sequence encoder learns user content preferences, and a mix-modality sequence decoder grasps the intrinsic relationship between these two types of preferences. Finally, we aggregate these three outputs as the fine-grained user preferences.

\subsubsection{Embedding Layer}
We construct an item-ID embedding matrix $E_{id}\in \mathbb{R}^{|I|\times d}$, where $d$ is the latent dimension of the vector. Given a user behavior sequence $S_{id}=\{i_1, i_2,...,i_n\}$, we look up the item-ID embedding matrix and add a learnable position embedding matrix $P\in \mathbb{R}^{n\times d}$ to obtain an input item-ID embedding matrix, 
\begin{equation}
\begin{aligned}
     M_e &= \text{Emb}(S_{id}) \\
      &= \left\{e_1^{id} + p_1, \ldots, e_j^{id} + p_j, \ldots, e_n^{id} + p_n\right\},
\end{aligned}
\end{equation}
where $e_j^{id} \in \mathbb{R}^{1\times d}$ and $p_j^{id} \in \mathbb{R}^{1\times d}$ are the corresponding $j$-th item-ID embedding vector and position embedding vector.

We then utilize the fine-tuned text encoder (i.e., BERT \cite{devlin-etal-2019-bert}) and image encoder (i.e., Swin-base \cite{liu2021Swin}) in Section \ref{item_level_modality_alingment}, to obtain an item-text embedding matrix $E_{text}\in \mathbb{R}^{|I|\times d_{text}}$, and an item-image embedding matrix $E_{ima}\in \mathbb{R}^{|I|\times d_{ima}}$, where $d_{text}$ and $d_{ima}$ are the output dimensions. Given an item-text sequence $S_{text} = \{t_1,t_2,...,t_n\}$ and an item-image sequence $ S_{ima}= \{g_1, g_2, ..., g_n\}$, we firstly encode them by looking up the corresponding embedding matrix and then transform their dimension to match the ID embedding via an adapter layer, which is implemented by a multi-layer perceptron layer. Secondly, we add the shared learnable position embedding matrix $P\in \mathbb{R}^{n\times d}$ for content input embedding matrix to learn temporal dependency,
\begin{equation}
\begin{aligned}
M_t &= Adapter(Emb(S_{text})) \\
    &= \{e_1^{text} + p_1, \ldots, e_n^{text} + p_n\} \label{eq:m_t},
\end{aligned}
\end{equation}
\begin{equation}
\begin{aligned}
M_g &= Adapter(Emb(S_{ima})) \\
  &= \{e_1^{ima} + p_1, \ldots, e_n^{ima} + p_n\},\label{eq:m_g}
\end{aligned}
\end{equation}
where $e_j^{text} \in \mathbb{R}^{1\times d}$ and $e_j^{image} \in \mathbb{R}^{1\times d}$ are the corresponding $j$-th text embedding vector and image embedding vector.

\subsubsection{ID-modality Sequence Encoder}\label{ID-encoder}
The user behavior interaction sequences contain important user preference signals. We apply the directional Transformer model SASRec \cite{sasrec} as an ID sequence encoder, which can be replaced by some other ID-based sequential models. Given the input item-ID embedding matrix $M_e\in \mathbb{R}^{n\times d}$, we first transform it into query, key, and value by linear projections, and then feed them to a self-attention module \cite{attention}, 
\begin{equation}
    \label{eq:att_all}
    \begin{aligned}
    \hat{H}_e &=Attention(M_e,M_e,M_e) \\
       &=Softmax\left(\frac{(M_eW^Q_e) ({M_eW^K_e})^T}{\sqrt d}\right)(M_eW^V_e),
    \end{aligned}
\end{equation}
where $W^Q_e, W^K_e, W^V_e \in \mathbb{R}^{d \times d}$ are learnable linear projection matrices. After that, we conduct a point-wise feed-forward network to get the hidden behavior matrix,
\begin{equation}
    \label{eq:att}
    \begin{aligned}
    H_e&=FFN(\hat{H}_e)\\
    &=ReLU(\hat{H_e}W_1^e+b_1^e)W_2^e+b_2^e,
    \end{aligned}
\end{equation}
where $W_1^e, W_2^e \in \mathbb{R}^{d \times d}$ are learnable weight matrices, and $b_1^e, b_2^e \in \mathbb{R}^{1 \times d}$ are learnable bias vectors. $H_e=\{h_1^e,h_2^e,...,h_n^e\} \in \mathbb{R}^{n \times d} $ is the output hidden behavior preference sequence. We use the last-step output $h_n^e$ as the user behavior preferences. 

\subsubsection{Content-modality Sequence Encoder} \label{con-encoder}
The item content information like text and image contains rich semantic signals. We aim to capture a user's content-based preferences from the item-content sequence. Given an item-text embedding matrix $M_t \in \mathbb{R}^{n\times d}$ and an item-image embedding matrix $M_g \in \mathbb{R}^{n\times d}$, we use an L2 normalization to combine the item-text and item-image representations in each position as a unified content one \cite{unsper}, 
\begin{equation}
e_i^{cont} = \frac{e_i^{text} + e_i^{ima}}{\|e_i^{text} + e_i^{ima}\|}.
\end{equation}
Therefore, we obtain a unified content embedding sequence $M_c=\{e_1^{cont},e_2^{cont},...,e_n^{cont}\} \in \mathbb{R}^{n \times d}$. The L2 normalization term has several advantages. Firstly, the unified content representations of the text modality and image modality improve rich semantic understanding and address the limitations inherent in uni-modal representations. For example, when the image modality lacks contextual information, the text modality effectively supplements it with semantic content. Secondly, it only needs a content encoder to model unified content representations rather than two individual encoders to model two types of content representations, thereby reducing the complexity of the preference model.

Then, similar to the ID-modality sequence encoder, we utilize a directional Transformer model to learn the content preferences. Firstly, we feed the content input matrix $M_c$ into  a self-attention module, 
\begin{equation}
    \label{eq:att_all}
    \begin{aligned}
    \hat{H}_c &=Attention(M_c,M_c,M_c) \\
       &=Softmax\left(\frac{(M_cW^Q_c) ({M_cW^K_c})^T}{\sqrt d}\right)(M_cW^V_c),
    \end{aligned}
\end{equation}
where $W^Q_c, W^K_c, W^V_c \in \mathbb{R}^{d \times d}$ are learnable linear projection matrices. Then we feed $\hat{H}_c$ into the feed-forward network, 
\begin{equation}
\begin{aligned}
    H_c&=FFN(\hat{H}_c) \\
    &=ReLU(\hat{H_c}W_1^c+b_1^c)W_2^c+b_2^c,
\end{aligned}
\end{equation}
where $W_1^c, W_2^c \in \mathbb{R}^{d \times d}$ are learnable weight matrices, and $b_1^c, b_2^c \in \mathbb{R}^{1 \times d}$ are learnable bias vectors. $H_c=\{h_1^c,h_2^c,...,h_n^c\} \in \mathbb{R}^{n \times d}$ is the output hidden content preference sequence. We use the last-step output $h_n^c$ as the user content preferences. 

\subsubsection{Mix-modality Sequence Decoder}
Previous encoders often independently model the ID sequences and content sequences and do not consider their inherent correlations. Therefore, we introduce a mix-modality sequence decoder to grasp the ID modality and content modality relations for learning better sequence representations. 

Given a behavior preference sequence  $H_e \in \mathbb{R}^{n \times d} $ from Section ~\ref{ID-encoder} and a content preference sequence $H_c \in \mathbb{R}^{n \times d}$ from Section ~\ref{con-encoder}, we adopt
$H_e$ as the query, and $H_c$ as the key and value in the cross-attention mechanism. The attention map in cross-attention captures the relationship between the behavior preferences $H_e$ and the content preferences $H_c$. Then, this attention map would be combined with the original content preferences $H_c$ to generate the output $\hat{H_m}$. The formulation is as follows,  
\begin{equation}
\begin{aligned}
    \hat{H_m}&=Attention(H_e,H_c,H_c) \\
    &=Softmax\left(\frac{(H_eW^Q_m) ({H_cW^K_m})^T}{\sqrt d}\right)(H_cW^V_m),
\end{aligned}
\end{equation}
where $W^Q_m, W^K_m, W^V_m \in \mathbb{R}^{d \times d}$ are learnable linear projection matrices. Similarly, we feed $\hat{H_m}$ into a feed-forward layer to obtain the mix-modality preferences,
\begin{equation}
\begin{aligned}
    H_m&=FFN(\hat{H}_m) \\
    &=ReLU(\hat{H_c}W_1^m+b_1^m)W_2^m+b_2^m,
\end{aligned}
\end{equation}
where $W_1^m, W_2^m \in \mathbb{R}^{d \times d}$ are learnable weight matrices, and $b_1^m, b_2^m \in \mathbb{R}^{1 \times d}$ are learnable bias vectors. $H_m=\{h_1^m,h_2^m,...,h_n^m\} \in \mathbb{R}^{n \times d} $ is the output hidden preference sequence. We use the last-step output $h_n^m$ as the mix-modality preferences. 

\subsubsection{Preference Aggregation} 
We aggregate the behavior preferences, content preferences, and mix-modality preferences, and obtain the final preference representations by a linear projection, 
\begin{equation}
  h_n^a=Projection(Concat(h_n^e,h_n^c ,h_n^m)),
\end{equation}
where $h_n^a \in \mathbb{R}^{1 \times d} $ is the user's final sequential preferences. 

\subsection{A Two-step Training Strategy} \label{training_strategy}
Training a sequence model from scratch is often inefficient and costly, because the content representations and ID representations may interfere with each other, leading to difficulty in model convergence and unstable performance. Therefore, we propose a two-step training strategy, i.e., (i) pre-train the ID modality sequence encoder with a standard cross-entropy loss and fix its weight; and (ii) post-train the content sequence encoder and mix-modality sequence decoder. We propose a content-aware contrastive learning to align the content modality representations and the ID modality representations, and utilize a low-rank adaptation layer to extract the user behavior preferences. In this setting, we decouple the training process of the content-modality dependency and the item-collaborative dependency. 

\subsubsection{Next-item Prediction Task}
In the first step, we utilize the next-item prediction task to train an ID-modality encoder and fix its weights in the second step. 

The prediction score for item $i$ as the next preferred item can be estimated as,
\begin{equation}
    \hat{y_i} = h_n^a (e_i^{id})^T,
\end{equation}
where $\hat{y_i}$ is the predicted score of item $i$ and $e_i^{id}$ is the $i$-th item-ID embedding vector. We adopt the cross-entropy loss function to measure the difference between the prediction $\hat{y}$ and the ground truth $y$, 
\begin{equation}
    \mathcal{L}_{CE} = -\sum_{i=1}^{|I|} y_i \log(\widehat{y}_i).
\end{equation}

\subsubsection{Content-aware Contrastive Learning Task}
We introduce a content-aware contrastive learning task to align user preferences with content modality representations for sequence-level modality representation alignment. Given a batch of $\textit{B}$ training instances $\{<h_1^a,e_1^c>,<h_2^a,e_2^c>,...,<h_B^a,e_B^c> \}$, where the $j$-th training instance is a pair of aggregated sequential preference representations $h_j^a$ and content representations $e_j^c$. We define the contrastive learning loss as follows,
\begin{equation}
    \mathcal{L}_{ConCL} = -\frac{1}{B} \sum_{i=1}^{B} \log \left( \frac{\exp( sim(h_i^a, e_i^c) / \tau)}{\sum_{i'=1}^{B} \exp( sim(h_i^a, e_{i'}^c) / \tau)} \right),
\end{equation}
where in-batch negative instances $\{e_{i'}^c\}$ are the content embeddings of the positive items of another sequences and $\tau \in[0,1]$ is the temperature parameter. 

\subsubsection{Low-rank Adaptation Layer}
The final loss function can be formulated as,
\begin{equation}\label{loss}
    \mathcal{L} = \mathcal{L}_{CE} + \alpha \mathcal{L}_{ConCL} + \lambda \|\mathbf{\Theta}\|_F,
\end{equation}
where $\mathbf{\Theta} = \{ E_{id}, E_{text}, E_{image}\}$ and $\| \cdot \|_F$ denotes the L2 normalization. Note that $ \alpha \in [0,1]$ is the balance parameter and $\lambda$ is 
the regularization parameter. 

Inspired by the low-rank adaptation method \cite{lora}, we adopt a LoRA layer to fine-tune the ID-modality sequence encoder and post-train other components with the final loss $\mathcal{L}$ in end-to-end training.

Given the fixed parameter weight $W_0 \in R^{d \times k}$ of the ID-modality sequence encoder, we represent its updating process as follows,
\begin{equation}
    W_0 + \Delta W = W_0 + BA,
\end{equation}
where $\Delta W=BA$ is the updating weight, and $B \in R^{d \times r}, A \in R^{r \times k}$
are learnable lightweight matrices. Note that $r << min (d, k)$ is the rank of the matrix. We initialize matrix $A$ with a zero-mean normal distribution and matrix $B$ with zeros.

\subsection{Analysis of Time Complexity}
In content modality semantic alignment, we use LLM-driven semantic discriminator to select a sample data, and fine-tune the text encoder and image encoder jointly based on the sample data. Then, we encode the text embedding and the image embedding matrix via these tuned encoders, which can be cached in advance. Therefore, there is no need to invoke LLM and content encoders in sequence preference learning, which reduces the extra inference costs.

In sequence preference learning, we adopt a two-step training strategy to train our model. In the prediction stage, we remove the content-aware contrastive learning task and infer user preferences based on the encoder-decoder model.

We assume each LLM inference takes time $t$ and there are $k$ samples. We show the time complexity of our SICSRec in Table~\ref{Time_Complexity}, where $b$, $n$, and $d$ denote the batch size, input sequence length, and hidden dimension, respectively. Our SICSRec has the same level of time complexity with Transformer-based sequential recommendation methods.

\begin{table}[h]
\centering
\caption{Time complexity analysis of our SICSRec.}
\label{Time_Complexity}
\begin{tabular}{ccc} 
\toprule
Component & \multicolumn{1}{c}{Time Complexity} \\
\midrule
LLM inference & \(O(t \cdot k ) \) \\
$\mathcal{L}_{SFT}$ & \(O(b^2 \cdot d ) \) \\ 
IMSE & \(O(b \cdot (n^2 \cdot d + n \cdot d^2) \) \\
CMSE & \(O(b \cdot (n^2 \cdot d + n \cdot d^2) \) \\
MMSE & \(O(b \cdot (n^2 \cdot d + n \cdot d^2) \) \\
$\mathcal{L}_{ConCL}$ & \(O(b^2 \cdot d ) \) \\ 
Training & \( O(b \cdot (n^2 \cdot d + n \cdot d^2) +b^2 \cdot d) \) \\
Inference & \( O(b \cdot (n^2 \cdot d + n \cdot d^2)) \) \\
\bottomrule
\end{tabular}
\label{tab:complexity}
\end{table}

\section{Experiments} 
In this section, we conduct extensive experiments to answer the following five research questions:
\begin{itemize}
\item \textbf{RQ1}: How does our SICSRec perform against the state-of-the-art ID-based and content-based SR methods? (see Sec.~\ref{sec:overall_performance})
\item \textbf{RQ2:} What is the impact of supervised fine-tuning of the content encoder on our SICSRec? (see Sec.~\ref{sec:supervised fine-tuning})
\item \textbf{RQ3:} What are the effects of different components in our SICSRec? (see Sec.~\ref{sec:ablation_study})
\item \textbf{RQ4:} Does the two-stage training strategy contribute to preference learning in our SICSRec? (see Sec.~\ref{sec:training_strategy})
\item \textbf{RQ5:} How do the hyper-parameters affect the performance of our SICSRec? (see Sec.~\ref{sec:pa})
\item \textbf{RQ6:} How is the inference efficiency of our SICSRec? (see Sec.~\ref{sec:infer})
\item \textbf{RQ7:} How can we visually demonstrate the impact of model modifications in our SICSRec? (see Sec.~\ref{sec:case_study})
\end{itemize}

\subsection{Datasets and Evaluation Metrics}\label{dataset}
We use four datasets from NinRec ~\cite{ninrec}, which is a large-scale multi-modality benchmark collected from an online video platform BiliBili with different scenarios. Each item in these datasets is associated with an item ID, a piece of descriptive text, and a high-resolution cover image. We show the statistical details of the datasets in Table ~\ref{tab:datasets}. We adopt the leave-one-out strategy to split each dataset, which uses the last item for test, the penultimate one for validation, and the other items for training. We use the full-ranking strategy for a fair comparison \cite{metrics}. Two commonly used metrics, i.e., Hit$@$K and NDCG$@$K are used for evaluation, where $K\in \{5, 10\}$.

\begin{table}[t]
  \caption{Statistical details of the datasets.}
  \label{tab:datasets}
\begin{tabular}
{p{1.5cm}p{1.0cm}p{1.0cm}p{1.5cm}p{0.9cm}}
\toprule
Dataset & \#Users & \#Items & \#Interactions  &  Sparsity  \\ 
\midrule
 Cartoon &  30,300 &  4,724 & 215,443 &  99.88\% \\
Dance & 10,715 & 2,307 & 83,392 & 99.66\% \\
 Food & 6,549 & 1,579 & 39,740 & 99.62\% \\
 Movie & 16,525 & 3,509 & 115,576 & 99.80\% \\
  
\bottomrule
\end{tabular}
\end{table}

\begin{table*}[ht!]
\caption{Recommendation performance comparison of eleven baselines and our SICSRec on the four datasets. Bold scores represent the best performance, while underlined scores indicate the second-best performance. "\textit{T}", "\textit{V}" and "\textit{ID}" stand for text, image, and ID. "Improved" denotes the relative improvement of our SICSRec compared with the second-best method. 
}\label{tab:performance}
    \centering
    \resizebox{\textwidth}{!}{
    
       \begin{tabular}{lcccccccccccccc}
        \toprule
        \multicolumn{2}{c}{\textbf{Input Type \& Model $\rightarrow$}} & \multicolumn{3}{c}{\textbf{ID}} & \multicolumn{3}{c}{\textbf{T}} & \multicolumn{3}{c}{\textbf{T+ID}} & \multicolumn{3}{c}{\textbf{T+V+ID}} & \textbf{Improved} \\
        \cmidrule(lr){3-5} \cmidrule(lr){6-8} \cmidrule(lr){9-11} \cmidrule(lr){12-13} \cmidrule(lr){14-14}
        \textbf{Dataset} & \textbf{Metric} & \textbf{GRU4Rec} & \textbf{SASRec} & \textbf{BERT4Rec} & \textbf{UniSRec} & \textbf{VQ-Rec} & \textbf{MISSRec} & \textbf{LLMESR} & \textbf{UniSRec}  & \textbf{MISSRec} &\textbf{LLMESR} & \textbf{MISSRec} & \textbf{SICSRec} \\
        \midrule

         \multirow{4}{*}{Cartoon} 
        & Hit@5 & 0.0678 & 0.0845 & 0.0354 & 0.0754 & 0.0470 & 0.0465 & 0.0713 &0.0801 & \underline{0.0856} & 0.0716 & 0.0730 & \textbf{0.0875} &  2.22\% \\
        & Hit@10 & 0.1074 & 0.1318 & 0.0628 & 0.1293 & 0.0909 & 0.0683 & 0.1120 &\underline{0.1327} & 0.1296 & 0.1130 & 0.1166 & \textbf{0.1344} &  1.28\% \\
        & NDGC@5 & 0.0431 & \underline{0.0474} & 0.0213 & 0.0432 & 0.0280 & 0.0241 & 0.0400 &0.0464 & 0.0454 & 0.0415 & 0.0430 & \textbf{0.0531} &  12.03\% \\
        & NDCG@10 & 0.0558 & 0.0626 & 0.0300 & 0.0607 & 0.0421 & 0.0308 & 0.0531 &\underline{0.0634} & 0.0589 & 0.0548 & 0.0564 & \textbf{0.0682} &  7.57\% \\
        \midrule
        
        \multirow{4}{*}{Dance} 
        & Hit@5 & 0.1395 & 0.1489 & 0.0914 & 0.1387 & 0.1029 & 0.0914 & 0.1379 & 0.1420 & 0.1266 & 0.1392 & \underline{0.1535} & \textbf{0.1578} &  2.80\% \\
        & Hit@10 & 0.2067& \underline{0.2251} & 0.1495 & 0.2155 & 0.1670 & 0.1409 &0.2088 & 0.2150 & 0.1876 & 0.2102 & 0.2248 & \textbf{0.2295} &  1.95\% \\
        
        & NDGC@5 & 0.0928 & 0.0933 & 0.0583 & 0.0883 & 0.0660 & 0.0505 & 0.0877 & 0.0897 & 0.0696 &0.0887 & \underline{0.0965} & \textbf{0.1074} &  11.30\% \\
        
        & NDCG@10 & 0.1143 & 0.1179 & 0.0770 & 0.1131 & 0.0866 & 0.0657 & 0.1106 & 0.1133 & 0.0883 & 0.1115 & \underline{0.1184} & \textbf{0.1304} &  10.14\% \\
        \midrule
    \multirow{4}{*}{Food}
        & Hit@5 & 0.0866 & 0.1356 & 0.0437 & 0.0556 & 0.0779 & 0.0739 & 0.1244 & 0.0999 & 0.1283 & 0.1266 & \textbf{0.1448} & \underline{0.1367} & -5.59\% \\
        & Hit@10 & 0.1376 & 0.1993 & 0.0765 & 0.1026 & 0.1460 & 0.1222 & 0.1849 & 0.1587 & 0.1855 &  0.1881 & \underline{0.2023} & \textbf{0.2037} &  0.69\% \\
        & NDCG@5 & 0.0561 & 0.0758 & 0.0265 & 0.0332 & 0.0440 & 0.0388 & 0.0711 & 0.0582 & 0.0631 & 0.0719 & \underline{0.0837} & \textbf{0.0849} &  1.43\% \\
        & NDCG@10 & 0.0724 & 0.0964 & 0.0370 & 0.0481 & 0.0658 & 0.0535 & 0.0906 & 0.0771 & 0.0807 & 0.0917 & \underline{0.1013} & \textbf{0.1064} &  5.03\% \\
    \midrule
    
    \multirow{4}{*}{Movie}
        & Hit@5 & 0.0618 & \underline{0.0784} & 0.0378 & 0.0761 & 0.0545 & 0.0415 & 0.0694 & 0.0765 & 0.0687 & 0.0734 & 0.0764 & \textbf{0.0810} &  3.32\% \\
        
        & Hit@10 & 0.0966 & 0.1247 & 0.0623 & 0.1229 & 0.0953 & 0.0631 & 0.1102 & \underline{0.1248} & 0.1107 & 0.1105 & 0.1155 & \textbf{0.1260} &  0.96\% \\
        
        & NDCG@5 & 0.0404 & 0.0434 & 0.0226 & 0.0445 & 0.0332 & 0.0236 & 0.0383 & \underline{0.0461} & 0.0369 & 0.0412 & 0.0446 & \textbf{0.0495} &  7.38\% \\
        
        & NDCG@10 & 0.0516 & 0.0583 & 0.0304 & 0.0595 & 0.0464 & 0.0301 & 0.0515&\underline{0.0617} & 0.0498 & 0.0532 & 0.0565 & \textbf{0.0640} &  3.73\% \\
        \bottomrule
        \end{tabular}
    }
\end{table*}

\subsection{Baselines}\label{base}

\noindent(1) Item ID-based sequential recommenders
\begin{itemize}
\item GRU4Rec~\cite{GRU4Rec}: A session-based recommendation method, which models user behavior sequences via a GRU network.
\item SASRec~\cite{sasrec}: A self-attentive sequential recommendation method, which models behavior sequences via a left-right unidirectional Transformer.
\item BERT4Rec~\cite{bert4rec}: A bidirectional Transformer-based sequential recommender with a mask-item modeling task.
\end{itemize}

\noindent(2) Item content-based sequential recommenders
\begin{itemize}
\item UniSRec(T)~\cite{unsirec}: A text modality-based sequential recommendation method that utilizes multi-domain text to learn universal item representations. 
\item  VQ-Rec~\cite{vqrec}: A vector-quantized sequential recommendation method that designs text representations as multiple code representations. 
\item  MISSRec(T)~\cite{MISSRec}: A text-modality-based sequential recommendation method that utilizes text to model user intents. 
\end{itemize}

\noindent(3) Item ID and content-based sequential recommenders
\begin{itemize}
\item UniSRec(T+ID)~\cite{unsirec}: An improved version of UniSRec, which fine-tunes item embedding for downstream recommendation tasks. 

\item MISSRec(T+ID)~\cite{MISSRec}: An improved version of MISSRec, combining item embedding and text embedding for user preference modeling. 

\item LLMESR(T+ID) ~\cite{LLMESR}: A dual-view sequential recommendation method that combines semantic embeddings from a large language model and behavior embeddings from user-item interactions. 

\item LLMESR++(T+V+ID) ~\cite{LLMESR}: An improved version of LLMESR, which combines text embeddings and image embeddings as the input of the cross-attention mechanism. 

\item MISSRec(T+V+ID)~\cite{MISSRec}: An improved version of MISSRec, utilizing text, image, and item ID to model user preferences. 
\end{itemize}

\subsection{Implementation Details}\label{imple}
For a fair comparison, we implement our SICSRec and all the baselines by RecBole~\cite{recbole}. The codes of GRU4Rec, SASRec, and BERT4Rec come from the RecBole platform. Moreover, we use the public codes of UniSRec\footnote{https://github.com/RUCAIBox/UniSRec}, VQ-Rec\footnote{https://github.com/RUCAIBox/VQ-Rec}, LLMESR\footnote{https://github.com/liuqidong07/LLM-ESR} and MISSRec\footnote{https://github.com/gimpong/MM23-MISSRec}. We publish the datasets and source code of our SICSRec\footnote{https://github.com/donglinzhou/SICSRec}.
The latent dimension $d$ is tuned from \{64, 128, 256\}, and $d$=256 yields the best performance. Following the setting of previous works \cite{sasrec,bert4rec}, the maximum sequence length is 50. The temperature $\tau$ is 0.05 and the dropout rate is 0.5. The training batch size is 128 in Stage 1 and 1024 in Stage 2. We use early stopping with the patience of 10 epochs to prevent overfitting. The balance parameter $\alpha$ ranges from 0.1 to 1.0 with a step of 0.1. The regularization parameter $\lambda$ is chosen from \{0.001,0.0001\}. The rank $r$ is chosen from \{4, 8, 12, 16\}. We use the Adam optimizer with a learning rate of 1e-3. We search the hyperparameters of all the compared methods using validation data. 

For sample construction, we adopt different LLMs as semantic discriminators and compare their performance, including GLM-4-9B-Chat \cite{glm2024chatglm}, Hunyuan \cite{HUNYUAN}, Qwen1.5-14B-chat \cite{qwen}, and DeepSeek \cite{deepseek}. We use the text encoder from BERT \cite{devlin-etal-2019-bert}, RoBERTa \cite{roberta}, and Sentence-T5 \cite{sentencet5}. We use the image encoder from Swin-base \cite{liu2021Swin}, ViT \cite{vit}, and Resnet50 \cite{resnet}. The original text embedding size is 768 and the image embedding size is 1000. We conduct experiments on a Tesla V100-PCIe GPU with 32GB memory. 

\subsection{Overall Performance Comparison (RQ1)}\label{sec:overall_performance}
We report the experimental results in Table ~\ref{tab:performance} and have the following observations.
\begin{itemize}
\item Item ID-based sequential recommenders still achieve good performance on content-modality recommendation scenarios. For example, SASRec outperforms other baselines on Hit@5 on Movie and Hit@10 on Dance. GRU4Rec and BERT4Rec do not perform well. Previous works have also shown that the performance of BERT4Rec does not surpass SASRec under the full-ranking evaluation setting \cite{DIF,S3, ligtsasrec}.

\item Text modality-only sequential recommenders (i.e., VQ-Rec, UniSRec(T), and MISSRec(T) ) do not surpass the strong item-ID-based sequential recommenders (i.e., SASRec). Importantly, combining the text modality and ID modality can achieve a similar performance compared with the ID modality-based methods, and surpass their text modality-only versions, showing that ID modality is still an important feature in recommender systems. 

\item Jointly considering the image and text modalities always improves the recommendation performance. MISSRec (T+V+ID) achieves the second-best performance on Dance and Food and  UniSRec achieves the second-best performance on Cartoon and Movie, showing that using more content information enhances representation learning.
 
\item Our SICSRec achieves the best performance on almost all the datasets, with an average improvement of 8.04\% on NDCG$@$5 and 6.62\% on NDCG$@$10 compared with the best performing baseline. On the Food dataset, our SICSRec achieves the second-best performance on Hit@5 and remains superior on other metrics, likely due to its small scale. As shown in Table \ref{tab:datasets}, the Food dataset is relatively small, with only 6,549 users and 1,579 items. It is not difficult for a typical deep learning-based model to handle such a dataset. Therefore, the baseline models like MISSRec and SASRec achieve strong performance, while our SICSRec only shows marginal improvements. Moreover, using RoBERTa\&ViT as the modality encoder further improves the performance of our SICSRec on Food, which can be observed from Table \ref{tab:encoder}. 

\end{itemize}

\subsection{Supervised Fine-tuning Analyses (RQ2)}\label{sec:supervised fine-tuning}
\begin{figure}[h]
    \centering
    \includegraphics[width=1.0\columnwidth]{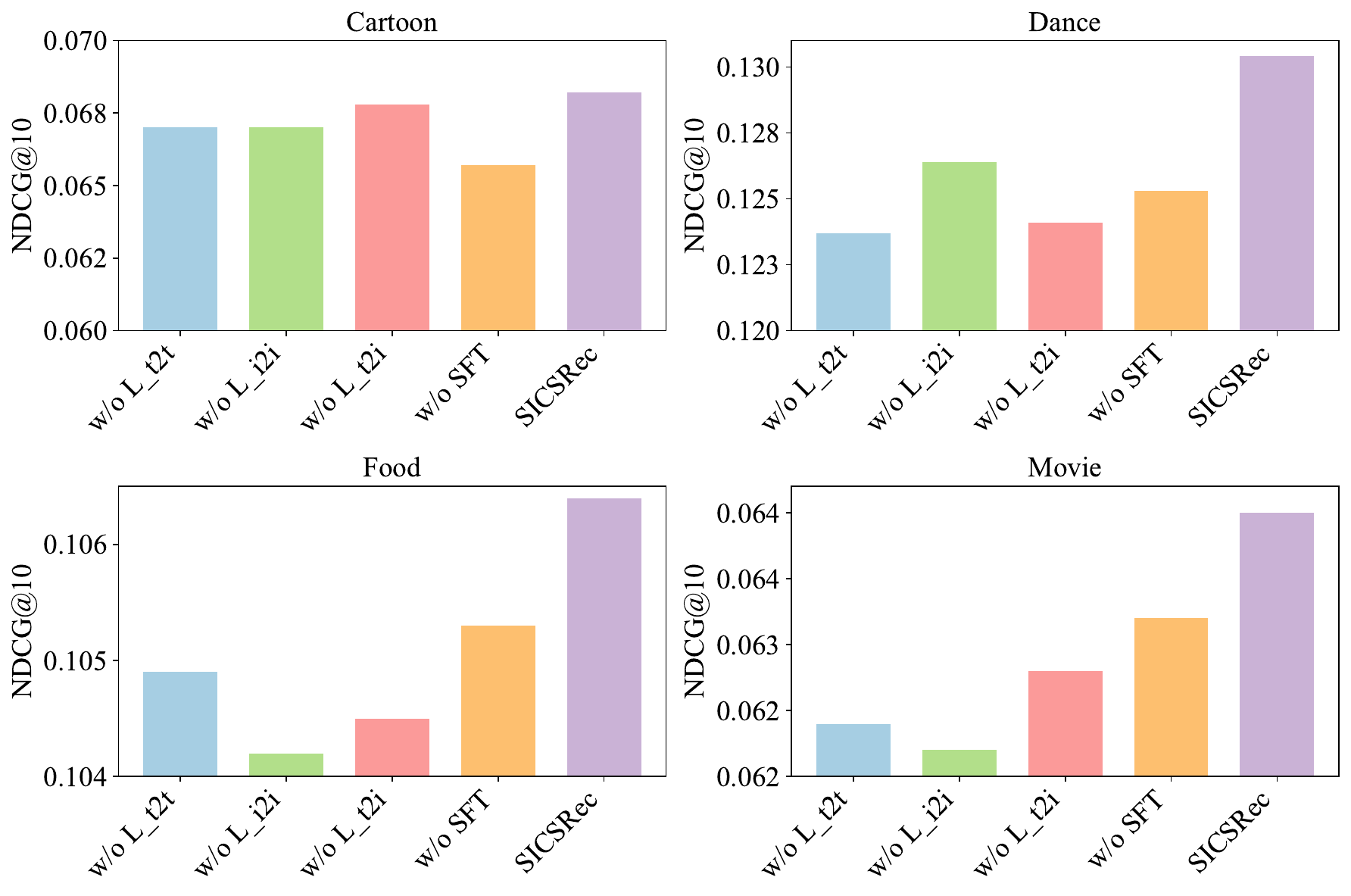}
    \caption{Supervised fine-tuning analyses. "t2t", "i2i", and "t2i" stand for text-to-text alignment, image-to-image alignment, and text-to-image alignment, respectively. "SFT" refers to the combination of these three alignment tasks.}
    \label{fig:sft}
\end{figure}

In this subsection, we examine the effectiveness of supervised fine-tuning and report the results in Figure ~\ref{fig:sft}. 

We have the following observations: (i) Supervised fine-tuning (SFT) of the text encoder and image encoder improves content representation learning and reduces the item-level semantic gaps. (ii) Removing any alignment task decreases the performance. (iii) The text-to-text alignment task plays a key role on Cartoon and Dance. Without the image-to-image alignment task, our SICSRec obtains the greatest drop in performance on Food and Movie.

Furthermore, we study using different LLMs as the semantic discriminator and compare their performance. The results are shown in Table ~\ref{tab: discriminator}. Hunyuan achieves the best Hit@10 performance on Dance while GLM achieves the best performance on Cartoon and Movie. Qwen gains the best performance on Food. DeepSeek also achieves comparable performance to that of GLM on most datasets. Different large language models will generate different samples, which may influence the fine-tuned content encoder, but their performance remains similar.

\begin{table}[thbp]
    \centering
    \renewcommand{\arraystretch}{1.2} 
    \caption{Recommendation performance with different LLMs as semantic discriminators. H@10 and N@10 stand for Hit@10 and NDCG@10, respectively.}
    \label{tab: discriminator}
    \resizebox{\linewidth}{!}{ 
    \begin{tabular}{l|cc|cc|cc|cc}
        \hline
        \multirow{2}{*}{Combination} & \multicolumn{2}{c|}{Cartoonn} & \multicolumn{2}{c|}{Dance} & \multicolumn{2}{c|}{Food}& \multicolumn{2}{c}{Movie} \\
         \cmidrule(lr){2-9} 
        &  H@10 & N@10 &  H@10 & N@10&  H@10 & N@10 &  H@10 & N@10 \\
        \hline 
       Hunyuan & {0.1337} & {0.0676} & \textbf{0.2324} & {0.1303} & {0.2011} & {0.1058} & {0.1254} & {0.0638}\\

        Qwen & {0.1343} & {0.0674} & {0.2316} & {0.1290} & \textbf{0.2054} & \textbf{0.1077} & {0.1253} & {0.0638}  \\
        
        DeepSeek & {0.1340} & {0.0679} & {0.2301} & {0.1284} & {0.2032} & {0.1070}  & {0.1248} & {0.0638}  \\
        
        GLM & \textbf{0.1344} & \textbf{0.0682} & {0.2295} & \textbf{0.1304} & {0.2037} & {0.1064} & \textbf{0.1260} & \textbf{0.0640}        \\
        \hline
    \end{tabular}
    } 
\end{table}

Finally, we study how different combinations of a text encoder and an image encoder affect the recommendation performance. Table \ref{tab:encoder} shows that the combination of BERT and Swin outperforms others on most datasets. RoBERTa\&ViT achieves the best NDCG@10 performance on Food and Movie, while Sentence-T5\&Resnet50 gains the worst results, which indicates that the combination of content encoders has a significant impact on the recommendation performance. We fix the text encoder (i.e., BERT) and replace the visual encoders. The results show that Swin performs better Hit@10 on Cartoon and Food, while ViT achieves higher performance on Dance and Movie. Overall, Transformer-based visual encoders (e.g., Swin and ViT) exhibit superior representation ability compared with ResNet-based architectures. Swin introduces a shifted window mechanism, enabling hierarchical feature extraction beyond ViT, so it may perform well in some scenarios. We also fix the visual encoder (i.e., Swin), and replace the text encoders. We find that BERT and RoBERT achieve comparable performance on most datasets, while Sentence-T5 gains lower results. This may be because encoder-only text encoders (e.g., BERT and RoBERTa) are generally more suitable for semantic understanding tasks than encoder-decoder models (e.g., Sentence-T5).
 
\begin{table}[thbp]
    \centering
    \renewcommand{\arraystretch}{1.2} 
    \caption{Recommendation performance with different combinations of text and image encoders. H@10 and N@10 stand for Hit@10 and NDCG@10, respectively.}
    \label{tab:encoder}
    \resizebox{\linewidth}{!}{ 
    \begin{tabular}{l|cc|cc|cc|cc}
        \hline
        \multirow{2}{*}{Combination} & \multicolumn{2}{c|}{Cartoonn} & \multicolumn{2}{c|}{Dance} & \multicolumn{2}{c|}{Food}& \multicolumn{2}{c}{Movie} \\
         \cmidrule(lr){2-9} 
        &  H@10 & N@10 &  H@10 & N@10&  H@10 & N@10 &  H@10 & N@10 \\
        \hline 
        BERT\&Swin  & \textbf{0.1344} & \textbf{0.0682} &0.2295 & \textbf{0.1304} & 0.2037 & 0.1064 & \textbf{0.1260}	& 0.0640 \\
        
        RoBERTa\&ViT  & 0.1324 & 0.0670 & \textbf{0.2304} &	0.1283	& \textbf{0.2042}	& \textbf{0.1076} & 0.1256 & \textbf{0.0643} \\
        
        Sentence-T5\&Resnet50    &  0.1322 & 0.0674 & 0.2285 & 0.1252 &	0.2014 & 0.1061 & 0.1245 & 0.0634 \\
        \hline
         BERT\&Swin  & \textbf{0.1344} & \textbf{0.0682} &	0.2295 & \textbf{0.1304} & \textbf{0.2037} & 0.1064 & 0.1260	& \textbf{0.0640} \\
         
        BERT\&ViT  & 0.1336  &	0.0671 & \textbf{0.2315} & 0.1283 &	0.2019 & \textbf{0.1071} & \textbf{0.1264} & 0.0638 \\
        
        BERT\&ResNet50  & 0.1317  &	0.0664 & 0.2249 & 0.1256 &	0.1987 & 0.1052 & 0.1241 & 0.0635 \\
        
        \hline
         BERT\&Swin  & \textbf{0.1344} & \textbf{0.0682} &	0.2295 & \textbf{0.1304} & \textbf{0.2037} & 0.1064 & \textbf{0.1260}	& \textbf{0.0640} \\
         
        RoBERTa\&Swin  & 0.1325 & 0.0674 & 0.2301 & 0.1283 & 0.2032 & \textbf{0.1077} & 0.1254 & 0.0638 \\
        
        Sentence-T5\&Swin  & 0.1321  &	0.0679 & \textbf{0.2305} & 0.1280 &	0.2016 & 0.1071 & 0.1250 & 0.0638 \\
        
        \hline
    \end{tabular}
    } 
\end{table}

\subsection{Ablation Study (RQ3)}\label{sec:ablation_study}
We study how each component affects the recommendation performance and report the ablation study results in Table ~\ref{tab:ablation}. Specifically, we study the contributions of the ID-modality sequence encoder (IMSE), the content-modality sequence encoder (CMSE), the mix-modality sequence decoder (MMSD), content-aware contrastive learning ($\mathcal{L}_{ConCL}$), the LoRA layer, and the L2-norm for CMSE.

\begin{table}[ht]
    \centering
    \renewcommand{\arraystretch}{1.0} 
    \caption{Recommendation performance (NDCG@10) in the ablation study.}
    \label{tab:ablation}
    \begin{tabular}{c|cccc}
        \hline
        Variants & Cartoon & Dance & Food & Movie \\
        \hline 
        w/o IMSE & 0.0622 &	0.1192 & 0.0977 & 0.0601 \\
        w/o CMSE & 0.0670 & 0.1260 & 0.1050	& 0.0625 \\
        w/o MMSD & 0.0665 & 0.1255 & 0.1046	& 0.0627 \\
        w/o $\mathcal{L}_{ConCL}$ & 0.0658 & 0.1222 & 0.1051	& 0.0604\\
        w/o LoRA layer &  0.0667 & 0.1246	& 0.1062 & 0.0617 \\
        w/o L2-norm & 0.0670 & 0.1252  & 0.1047 & 0.0634 \\
        Our SICSRec & \textbf{0.0682} & \textbf{0.1304} & \textbf{0.1064} & \textbf{0.0640}  \\
        \hline
    \end{tabular}
\end{table}
We have the following observations:
\begin{itemize}
\item The ID-modality sequence encoder is the most important component in our SICSRec. Removing it achieves the lowest performance on all datasets.
\item The content-modality sequence encoder and the mix-modality sequence decoder contribute to the performance of our SICSRec, because they learn the users' content preferences, and the inherent correlations between behavior preferences and content preferences.
\item Removing content-aware contrastive learning ($\mathcal{L}_{ConCL}$) decreases the performance because it contributes to the sequence-level representation alignment between user preferences and content representations.
\item Without the LoRA layer in the ID-modality sequence encoder, our SICSRec suffers from inadequate behavior preference learning.
\item We find that removing the L2 normalization term for CMSE and using the two encoders to model the text sequence and the image sequence would decrease the performance because the content modality gap may interfere with representation learning.
\end{itemize}

\begin{table}[ht]
\centering
 \renewcommand{\arraystretch}{1.0} 
\caption{Recommendation performance (NDCG@10) with different modality information.}
\label{tab:multi_input}
\resizebox{\linewidth}{!}{
\begin{tabular}{cccccc}

\toprule
Input Type & Variant & Cartoon & Dance & Food & Movie \\
\midrule
T    & UniSRec & \textbf{0.0607}	& 0.1131& 	0.0481	& \textbf{0.0595}
  \\
     & MISSRec & 0.0308	& 0.0657 & 0.0535& 0.0301 \\
    & SICSRec & 0.0591	& \textbf{0.1167} & \textbf{0.0944} &	0.0546  \\
\hline
V   & MISSRec & 0.0378	& 0.0931	& 0.0815 &	0.0429
  \\
    & SICSRec & \textbf{0.0621} & \textbf{0.1216}	& \textbf{0.0979} & \textbf{0.0592} \\
\hline
T+V & MISSRec & 0.0451 & 0.0983	& 0.0882 & 0.0466 \\
    & SICSRec & \textbf{0.0653} & \textbf{0.1298} & \textbf{0.1092} &\textbf{0.0605} \\
\hline
ID+T  & UniSRec & 0.0634 & 0.1133 & 0.0771 & 0.0617 \\
     & MISSRec & 0.0589 & 0.0883	& 0.0807 & 0.0498 \\
    & SICSRec & \textbf{0.0662}	& \textbf{0.1217}	& \textbf{0.1010}	& \textbf{0.0624}  \\
\hline
ID+V & MISSRec & 0.0625	& 0.1029 & 0.0828 & 0.0487  \\
     & SICSRec & \textbf{0.0665}	& \textbf{0.1243} &	\textbf{0.1056} & \textbf{0.0634} \\
\hline
ID+T+V & MISSRec & 0.0564 & 0.1184	& 0.1013 & 0.0565 \\
    & SICSRec & \textbf{0.0682}	& \textbf{0.1304}	& \textbf{0.1064}	& \textbf{0.0640} \\
\bottomrule
\end{tabular}}
\end{table}

Secondly, we investigate the capability of our SICSRec and other content-based sequential models in leveraging multi-modal information. The results are reported in Table ~\ref{tab:multi_input}. 

We have the following observations: (i) Removing the ID modality and only modeling user content preferences results in the worst performance, demonstrating that the ID modality plays a significant role. (ii) Our SICSRec has a large advantage over MISSRec when only the T or V modality is used. The performance gap narrows when both the T and V modalities are utilized, which indicates that considering more content modality information always enhances the recommendation performance. (iii) Our SICSRec outperforms other content-modality models in most cases, showing that our SICSRec is robust and competitive.

\subsection{Training Strategy of SICSRec (RQ4)}\label{sec:training_strategy}
We compare our two-step training strategy with different training strategies. Table ~\ref{tab:lora} shows their performance and the number of epochs required for model convergence. Note that the number of epochs may vary due to different random seeds and hardware.

\begin{table}[h]
    \centering
    \renewcommand{\arraystretch}{1.0} 
    \caption{Recommendation performance with different training strategies. N@10 and \#epo stand for NDCG@10 and the number of epochs, respectively. "Fixed" means fixing the model weight of this component.}
    \label{tab:lora}
    \resizebox{\linewidth}{!}{ 
    \begin{tabular}{l|cc|cc|cc|cc}
        \hline
        \multirow{2}{*}{Strategy} & \multicolumn{2}{c|}{Cartoonn} & \multicolumn{2}{c|}{Dance} & \multicolumn{2}{c|}{Food}& \multicolumn{2}{c}{Movie} \\
         \cmidrule(lr){2-9} 
        &  N@10 & \#epo & N@10 & \#epo  & N@0 & \#epo  & N@10 & \#epo  \\
        \hline 
        Fixed IDEnc  & \textbf{0.0682} &	17 &	\textbf{0.1304} &	51 & \textbf{0.1064} &	17	& \textbf{0.0640}	& 26 \\
        Fixed IDEmb   & 0.0653 &	34	  & 0.1263 & 24	& 0.1056 & 18 &	0.0618 & 13  \\
        Fixed IDEmb\&Enc  & 0.0656 &	17	  & 0.1254 & 22 & 0.1059 &	17	& 0.0608 & 17 \\
        \hline
        Fixed ConEnc  & 0.0470 &	76	& 0.1075	& 60 & 0.0610 & 135 & 0.0485 &	86 \\
        Fixed ConEmb  & 0.0476 &	110	&	0.1164 & 104 &	0.0644	& 116	&	0.0457 &	99 \\
        Fixed ConEmb\&Enc  & 0.0496 &	132	& 0.1084 & 64 & 0.0605 & 121 & 0.0513	& 118 \\
        \hline
        End2end(not fixed) & 0.0449 &	75 &	0.1102	& 51	& 0.0586	& 79	&0.0473	& 97 \\
        \hline
    \end{tabular}
    } 
\end{table}

The observations are as follows. (i) Training our SICSRec from scratch (end-to-end training) results in low efficiency and unstable performance, because the content representations and the ID representations may interfere with each other, leading to difficulty in model convergence. (ii) Pre-training content-modality-related components firstly requires more training epochs and achieves sub-optimal performance. (iii) Pre-training ID modality-related components and post-training other components (i.e., our two-step representation learning) achieve stable performance and require fewer training epochs for model convergence.

\subsection{Parameter Analyses (RQ5)}\label{sec:pa}
\begin{figure}[h]
    \centering
    \begin{minipage}[t]{0.48\textwidth}
        \centering
        \includegraphics[width=\linewidth]{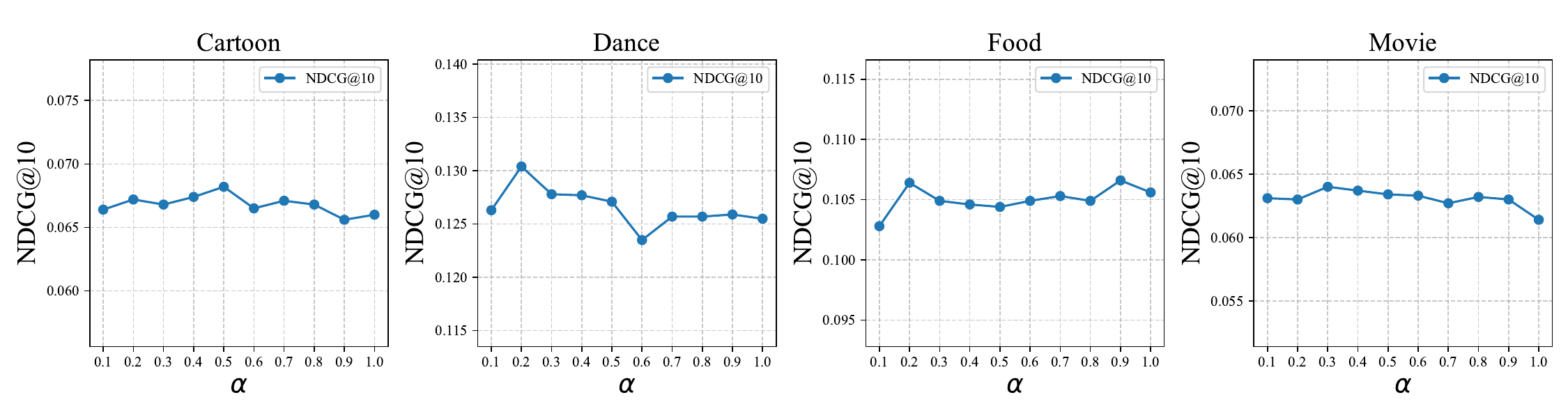}
        \vspace{0.5cm} 
        \includegraphics[width=\linewidth]{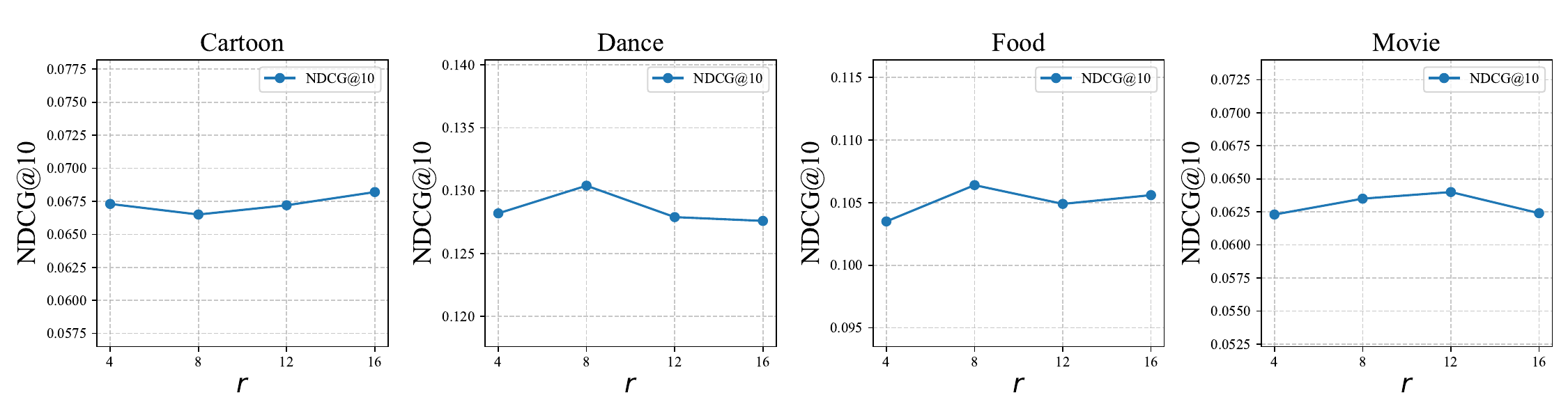}
        \caption{Recommendation performance of our SICSRec with different values of the balance parameter $\alpha$, and rank $r$.}
        \label{fig:parame}
    \end{minipage}
\end{figure}

In this experiment, we aim to investigate the effect of two hyper-parameters, i.e., the balance parameter $\alpha$ for $\mathcal{L}_{ConCL}$ and the LoRA rank $r$. Figure~\ref{fig:parame} shows that the contrastive learning ratio significantly impacts the performance of our SICSRec. Changing the value of $r$ does not greatly increase the performance, and setting a small rank for tuning the ID-modality sequence encoder is enough. 

\subsection{Inference Efficiency (RQ6)}\label{sec:infer}
\begin{figure}[htbp]
    \centering
    \includegraphics[width=1.0\linewidth]{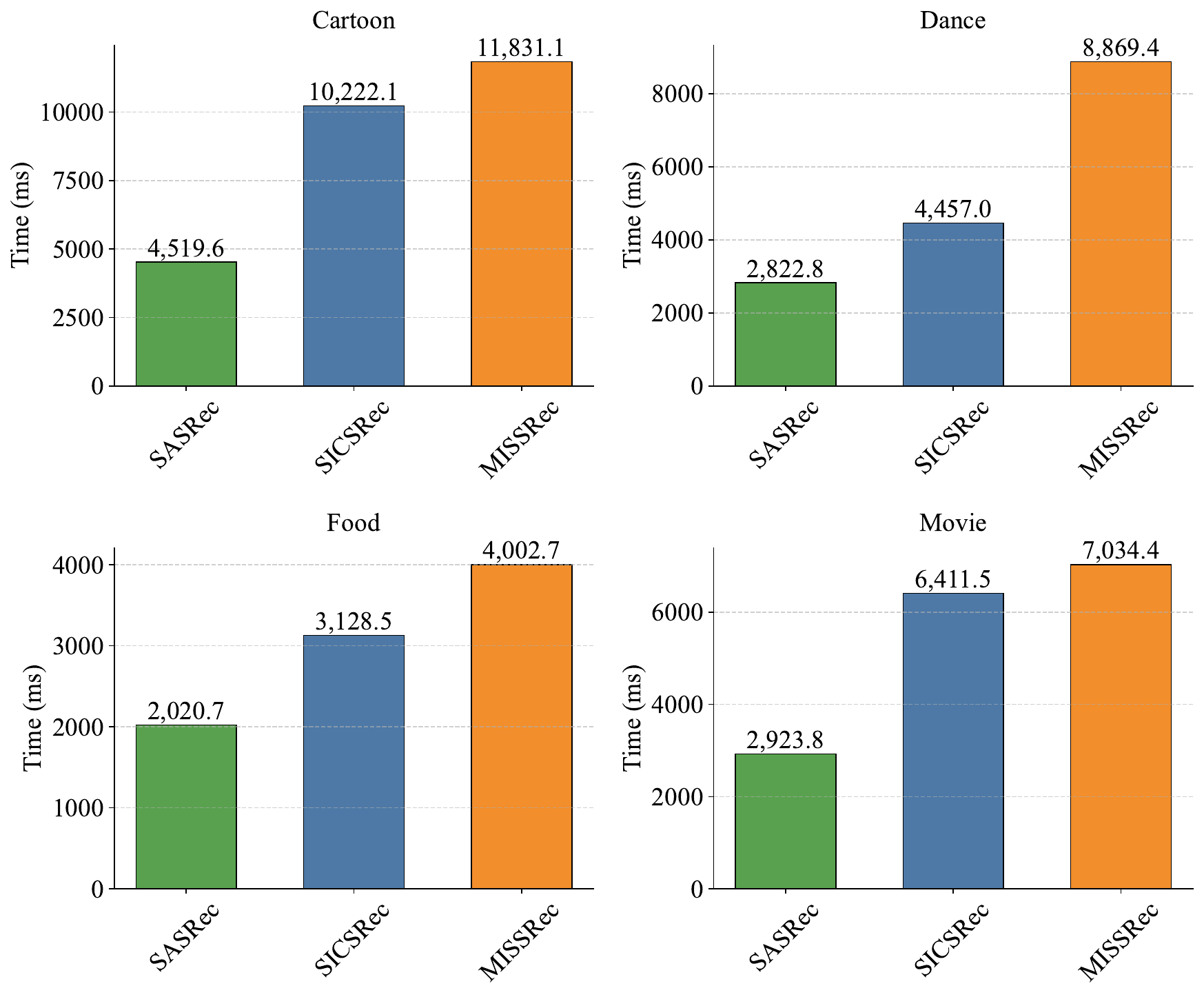}    
    \caption{Inference time of SASRec, MISSRec, and our SICSRec.}
    \label{fig:inference}
\end{figure}

We study the inference efficiency of SASRec, MISSRec, and our SICSRec. We conduct experiments on a single NVIDIA Tesla V100-PCIe GPU with 32 GB memory. We load the pre-trained model weights into GPU memory and measure the forward pass latency, excluding data loading and preprocessing time. Figure~\ref{fig:inference} shows that SASRec achieves the fastest inference speed, and our SICSRec demonstrates competitive efficiency by outperforming MISSRec. The inference efficiency is acceptable for real-world recommendation.

\subsection{Case Study (RQ7)}\label{sec:case_study}
We choose the Food scenario and conduct a case analysis on model modifications. We use the same user interaction sequence as input, load different pre-trained variant model weights to predict the next item, and compare the recommended item with the ground-truth. 

\begin{figure}[htbp]
    \centering
    \includegraphics[width=1.0\linewidth]{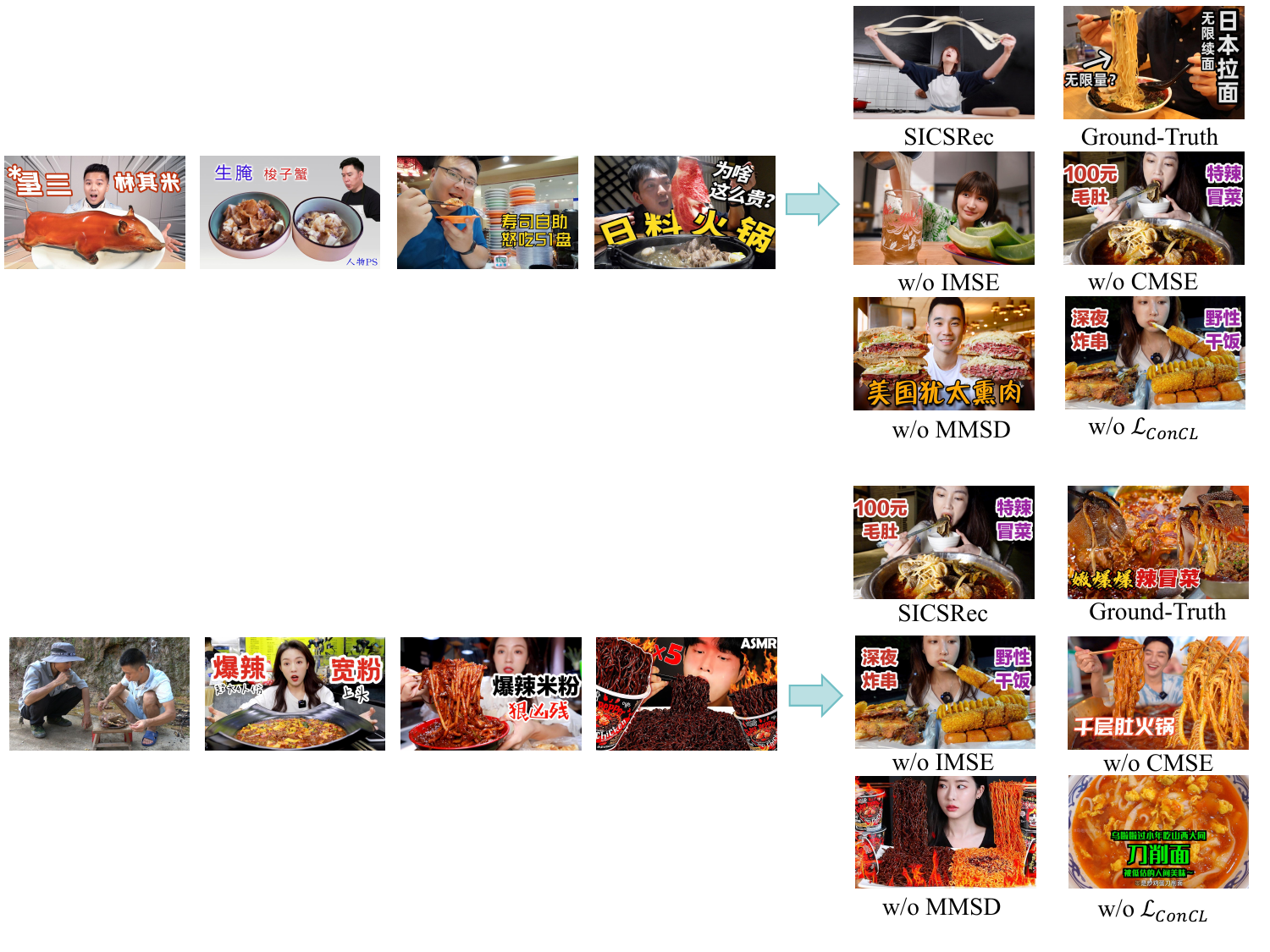}    
    \caption{Case studies of our SICSRec with different model modifications (left: historical preferred items; right: recommended items).}
    \label{fig:case}
\end{figure}

From Figure \ref{fig:case}, we can observe that our SICSRec recommends more accurate items than other variants by effectively combining visual, textual, and behavioral information. For example, in the first case, the user interaction history shows a stronger interest in Japanese food, and our SICSRec recommends a Japanese noodle dish, while other baseline models only recommend fast food.

\section{Conclusions and Future Work}
In this paper, we propose a novel self-supervised sequential representation learning method named SICSRec. Firstly, we propose a novel content modality semantic alignment module to reduce the item-level semantic gap between different modality representations. Then, we propose a novel Transformer-based encoder-decoder model to learn users' sequential preferences. Finally, we propose a two-step training strategy to train our model. We conduct extensive experiments and ablation studies to study the effectiveness of our SICSRec on four datasets and find that our model is very competitive compared with the state-of-the-art methods. 

In the future, we are interested in exploring some self-adaptive modality fusion methods for sequential recommendation with rich side information. Moreover, we plan to extend our solution to handle the cold-start problem with new users and new items.
 
\begin{acknowledgement}
We thank the support of Guangdong Basic and Applied Basic Research Foundation (No.2024A1515010122) and National Natural Science Foundation of China (Nos. 62461160311 and 62272315).
\end{acknowledgement}

\bibliographystyle{fcs}
\bibliography{ref}


\end{document}